\documentclass[a4paper]{article}
\usepackage{indentfirst}
\usepackage{multirow}
\usepackage{graphicx}
\usepackage{epsfig}
\usepackage{subfigure}
\usepackage{cite}
\usepackage{color}
\usepackage[top=1in, bottom=1in, left=1.25in, right=1.25in]{geometry}

\def\bea{\arraycolsep .1em \begin{eqnarray}}
\def\eea{\end{eqnarray}}

\def\s0#1#2{\mbox{\small{$ \frac{#1}{#2} $}}}
\def\0#1#2{\frac{#1}{#2}}

\begin{document}

\setcounter{topnumber}{10}
\setcounter{totalnumber}{50}
\title{The sinusoidal periodicity nature for $M\ge5$ global earthquakes}
\maketitle

\begin{center}
{\sc Z.~X.~Zhang,$^{\dagger\,,}$
\footnote{e-mail address:
zxzhang@neis.cn
 }\,\,\,
X.~Q.~Li$^{\star\,,}$\footnote{e-mail address:
lixq@ihep.ac.cn}}
\\
\vspace{0.5cm}

\noindent{\small{$^\dagger$ \it National Earthquake Infrastructure Service, Beijing, China}}\\
\noindent{\small{$^\star$ \it   Institute of High Energy Physics, Beijing, China}}
\end{center}
\begin{abstract}
By using the $M\ge5$ global earthquake data for Jan. 1950 to Dec. 2015, we performed statistical analyses for the parameters magnitude, time, and depth on a yearly scale. The magnitude spectrum, which is the earthquake number accumulated at different magnitudes, had an exponential distribution. For the first time, we report a very significant characteristic of the sinusoidal periodic variation in the spectral index. The cycle of the sine function fitting was 30.98 years. The concept of annual equivalent total magnitude (AETM) of total released energy for each year was introduced and the trend variation of AETM year by year was studied. Overall, the global AETM of earthquakes with $M\ge5$ displayed a certain upward trend as the years elapsed. At the same time, the change of the average epicenter depth of the global earthquakes ($M\ge5$) in each year was analyzed.

\end{abstract}

\section{Introduction}\label{intro}
As one of several planets that orbit around the Sun, Earth experiences many closely related motion cycles including the tropical year cycle, Earth's rotational cycle, lunar orbital periods and the precession cycle of the moon's orbit, and intersections with the zodiac. The research on earthquake periodicity dates back to several decades and variations are generally thought to be linked to the above cycles\cite{Schuster,Kasahara2002}. Notably, many studies have reported possible correlations and cyclic effects on earthquake processes resulting from solar and lunar tidal activities\cite{Cochran,Cadicheanu,Fidani2008, Metivier,Tanaka,Chen2012,Chao-Di}.

Rajesh and Tiwari\cite{Rajesh} used the methods of singular spectral analysis (SSA) and correlations to quantify the nature of principle dynamical processes that can affect the global annual earthquake rate. They point out that the earthquake rate has an 11 years cyclic variation corresponding to the well-known solar cycle.
 Studies on the cyclic characteristics of earthquake activities in the Yunnan province in China spanning 20 centuries by Hu {\it et al.}\cite{Hu} have showed that the node constituent caused by changes of obliquity in the lunar orbit is a possible mechanisms that can influence the long-term cyclic activities of earthquakes. They have also shown that earthquakes display a 11 year cyclic nature similar to the Sun's activity cycle.

  David {\it et al.}\cite{David2011} has presented evidence that the earthquake frequency in California is associated with cycles of 9 years and 56 years according to $M\ge6.9$ earthquake data analyzed over 200 years.
Du {\it et al.}\cite{Dupr2011} have reported that the precession of the orbit of the moon and the ecliptic plane intersection with an 18.6 year cycle may induce the tides by affecting the movement of the Earth's upper mantle, thus influencing the Earth's rotation and further triggering the occurrence of strong earthquakes. Ding {\it et al.}\cite{Ding1994} have shown that earthquake frequency is modulated by the lunar phase, which results in elevated earthquake magnitudes up to $25\%$. The periodic solar effect on earthquake activities as both a regional and global phenomenon has been reported by many other researchers\cite{Zhang1998, Odintsov2007,Jusoh2011,Tavares2011}.

In the 1960s, Simpson {\it et al.}\cite{Simpson} published a paper that discusses solar activity as trigger mechanism of earthquakes. Specifically he deduced the relationship between solar activity and earthquake activity by their similar cycle characteristics and pointed out the possibility that the occurrence of an earthquake can be triggered by sunspots that affect the ground electricity current. He also suggested that there was a negative correlation between seismicity and sunspot numbers. In 2006, Arcangelis {\it et al.}\cite{Arcangelis2006} compared the occurrence rule of sunspots and earthquakes and found that both of them obeyed a power exponential distribution in terms of the frequency and magnitude. Additionally, the duration distributions of aftershocks and small sunspots after large events both satisfy the Omori law. Hence they think there is a similar physics mechanism that drives the occurrence of sunspots and earthquakes. By the analysis of phase folding algorithms, Zhang {\it et al.}\cite{Zhang2013} studied the frequency characteristics of earthquakes associated with the solar and lunar movement cycles and found 11 regions modulated obviously by seasonal changes and 4 regions modulated by night -- day alternations. Qin {\it et al.}\cite{Qin2014} investigated the relation between solar activities and earthquakes and pointed out that the response variable is the frequency of earthquakes with Richter magnitudes of 4 -- 4.9.

There are also some models that simulate the periodicity of earthquake occurrences. Smith {\it et al.} \cite{Smith2006} performed simulations over a time span of 1000 years for the San Andreas fault by several kinds of cycle models of earthquake occurrence, which were based on the historical earthquake data and fitted by local geology features and plate parameters. Luttell {\it et al.}\cite{Luttrell2007} have shown that earthquake occurrence in the San Andreas region is related to stress changes in the lithosphere caused by water level variations of the Cahuilla River. Noa {\it et al.}\cite{Noa} simulated the high frequency earthquake activity and recreated the occurrence cycle of earthquakes, and the data were consistent with the historical record.

The earthquake belts in different areas have different geological structures which are generally related to the earthquake cycle. The earthquakes show cyclic characteristics in the areas of fault zones. However, the cycles in different areas are much different and most relevant studies of them focus on something local and regional.

The cycles of earthquake frequency cover several years to several decades and those of different earthquake belts display different features. In this paper, we examine the spectral index features of the earthquake magnitude distribution as a new type of analysis method. We regard the plates of the Earth as a whole body and perform macroscopic analyses of it. The global periodic characteristics all over the world reflect the influence on Earth from celestial bodies including the motion regulation of Earth itself as a whole body. By using the earthquake data with $M\ge5$ magnitudes from 1950 -- 2015 that were downloaded from the U.S Geological Survey (USGS), we analyzed the duration variation and periodic characteristics indicative of the magnitude proportion relation between large shocks and small shocks of annual earthquakes during an about 43 year period through the applications of new techniques.

\section{Data selection}
\begin{figure}
\centering
\includegraphics[width=\linewidth]{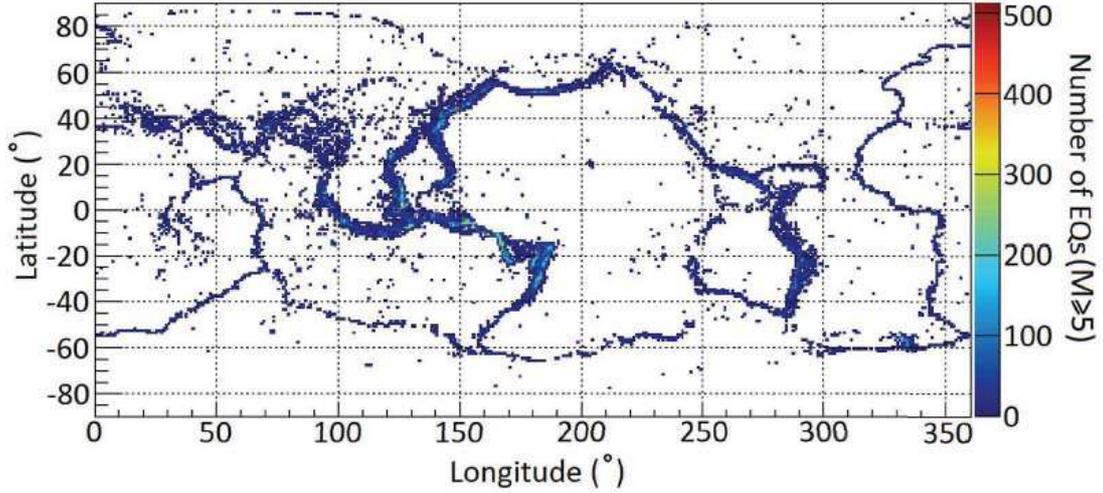}
  \caption{\small The earthquake distribution during Jan. 1950 to Dec. 2015, $M \ge 5$. }
  \label{1}
 \end{figure}

\begin{figure*}[t]
\includegraphics[width=12cm]{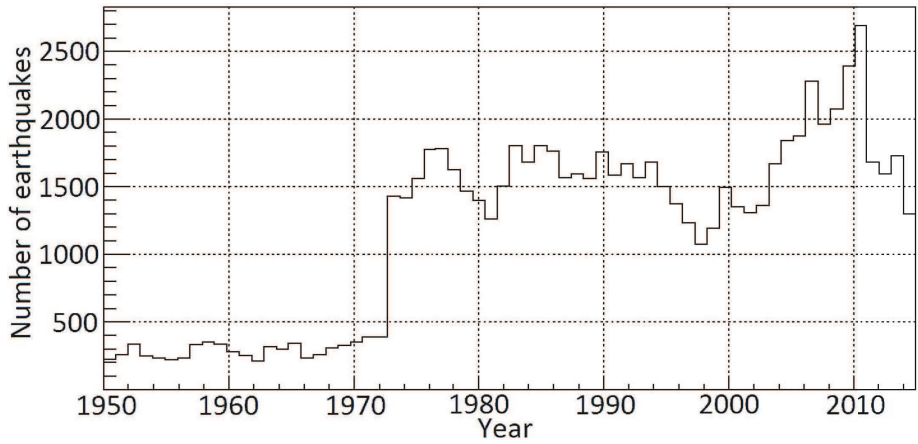}
\caption{The timing curve distribution during Jan. 1950 to Dec. 2015, $M \ge 5$.}
\label{2}
\end{figure*}

According to data points from earlier research\cite{Arcangelis2006}, the earthquake frequencies of different magnitudes obey the power exponential distribution, and in this paper, we define the earthquake magnitude spectrum, which indicates the occurrence frequency distribution of earthquakes corresponding to different magnitudes. The earthquake data used for the present study consisted of magnitude $M\ge5$ data for the period of January 1, 1950, to December 31, 2015, and the data were obtained from the USGS global catalog. The distributions of earthquakes and the occurrence frequency are shown in Fig.\ref{1} and the timing curve is presented in Fig.\ref{2}. From Fig.\ref{2}, we can see that the earthquake data for the period of January 1, 1950 to December 31, 1973 is not completely recorded.

\begin{figure*}[t]
\includegraphics[width=10cm]{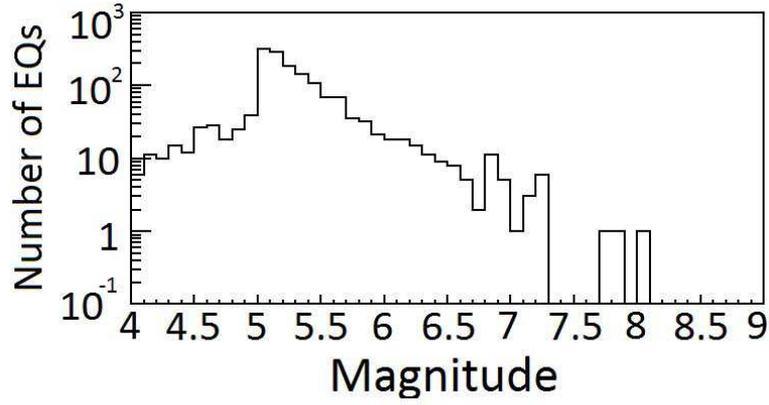}
\caption{The frequency distribution of different magnitude global earthquakes in 1980 with $M\ge4$ (magnitude spectrum).}
\label{3}
\end{figure*}

We did not adopt the seismic data for $M<5$ earthquakes because of the fragmentary characteristics of the database, as shown in the Fig.\ref{3}, which shows the earthquake magnitude spectrum with $M>4$ data, and this figure shows a sharp decline in the range of $M<5$, which is the reason that such data were not paid attention to in this work. In other words, to guarantee the completeness of the seismic data set, our analysis only focused on the earthquake data with $M\ge5$. Additionally, only seismic data after 1973 were considered for the credible analyses because these data were recorded by a more accurate detection method. Thus, both accuracy and completeness were a key data selection criteria used for the analyses of the earthquake magnitude spectrum.

\section{Analyses and results}
\subsection{Duration characteristics of the earthquake magnitude spectrum}

We can find from Fig.\ref{3} that, except for the incomplete data for lower magnitude, the earthquake magnitude spectrum with $M\ge5$ obeys the power exponential distribution.  The data for each year all over the world was accumulated into one spectrum which was fitted by the following exponential function:
\begin{equation}
F(M)=A\cdot exp(-\lambda\cdot M)
\end{equation}

where M indicates the magnitude of earthquakes, F denotes the earthquake frequency number for them corresponding to the M magnitude, A is the normalized factor that reflects the whole level of the frequency value, and $\lambda$ denotes the spectrum index that reflects the relative proportion of large and small earthquakes. Larger $\lambda$ values are indicative of more numerous small earthquakes.

\begin{figure*}[t]
\includegraphics[width=8cm]{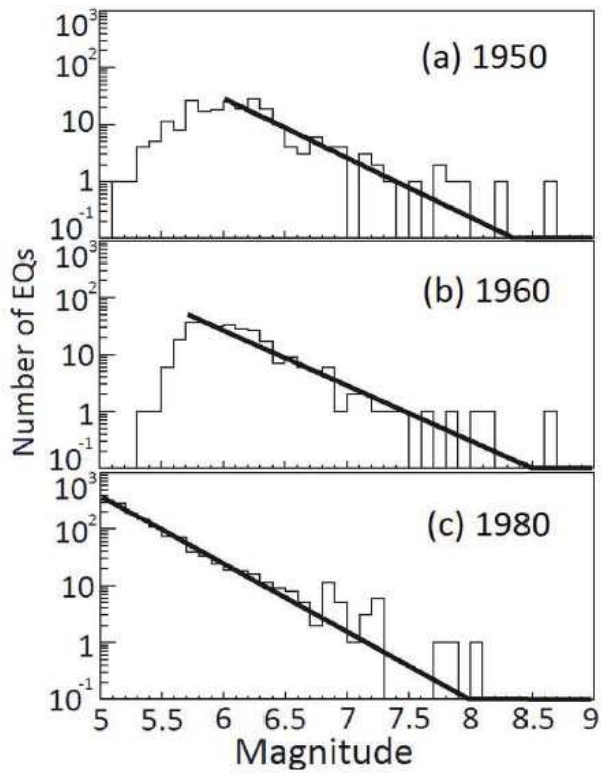}
\caption{The representative magnitude spectrum fittings for the global earthquakes in 1950, 1960 and 1980 respectively. Histogram: data, smooth line: fit result.}
\label{4}
\end{figure*}

The data range used to fit the magnitude spectrum is $M\ge6$ in 1950-1957, $M\ge5.7$ in 1958-1972 and $M\ge5$ in 1973-2015.
Figure \ref{4} shows the fitted result of the representative earthquake magnitude spectrum in 1950, 1960 and 1980 according to the exponential function (1).

\begin{figure*}[t]
\includegraphics[width=10cm]{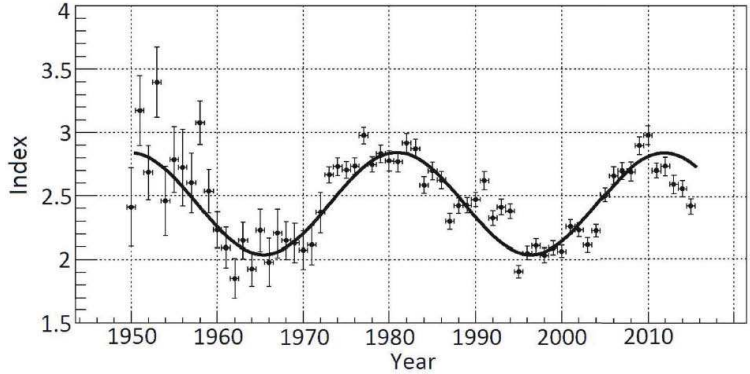}
\caption{The timing curve of the spectral indices and the fit result during Jan. 1950 to Dec. 2015, for $M\ge5$ earthquake data.
Spots with error: spectral indices, smooth line: fit result.
}
\label{5}
\end{figure*}

The fitted results of the magnitude spectrum in other years were similar with those of Fig.\ref{4}. By calculating annual statistics for the spectrum indices by fitting the magnitude spectrum for all earthquakes in this data during Jan. 1950 to Dec. 2015, we can obtain the index distribution with years, as shown in Fig.\ref{5}, which obviously obeys a sinusoidal function.

We structure a function coming from the sinusoidal function to fit the spectrum index distribution in Fig.\ref{5} as follows:
\begin{equation}
I(Y)=H+R\cdot sin(2\pi/C\cdot Y+P)
\end{equation}
Where I denotes the index value corresponding to the Y year, H is the constant factor, R is another constant factor indicating the amplitude of the sinusoidal function, and C and P denote the cycle and phase of the sinusoidal function respectively.

\begin{figure*}[t]
\includegraphics[width=10cm]{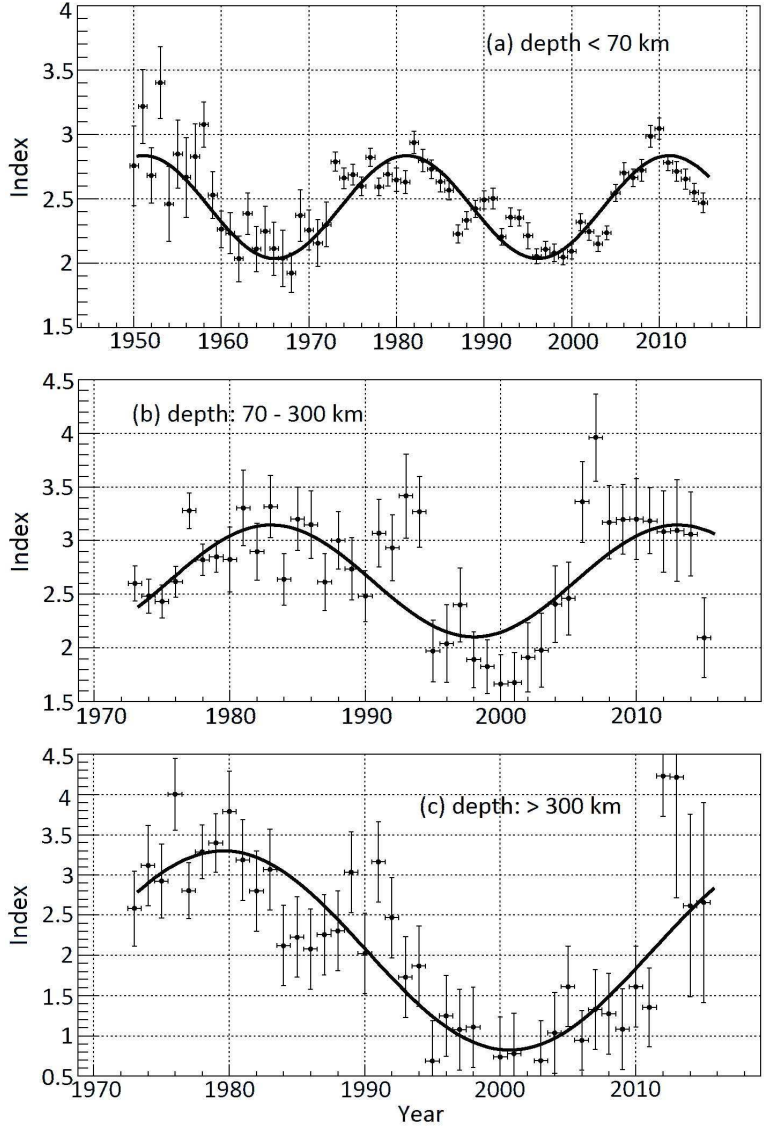}
\caption{The timing curve of the spectral indices at different depth ranges and the fit result for $M\ge5$ earthquake data.
Spots with error: spectral indices, smooth line: fit result.
}
\label{cycle}
\end{figure*}

 Because of the different mechanisms of earthquake occurrence for different epicenter depths, we analyzed the magnitude spectrum indices of data that were divided into three kinds of earthquakes including shallow-focus shocks below 70 km, intermediate-focus shocks at 70 -- 300 km, and deep-focus earthquakes over 300 km, as shown in Fig.\ref{cycle}. Due to the relatively fewer statistics for intermediate and deep-focus earthquakes in period of 1950-1972, the fitting data used in Fig.\ref{cycle} (b) and (c) only covers that from Jan. 1973 to Dec. 2015.

After fitting the spectrum indices with years by function (2) for the whole earthquake data set and the three data sets for different depths, the fitted parameters and the goodness of fit were calculated, and the results are shown in Table 1. The fitted parameter H in the structured sinusoidal function is equivalent to the average index value for each year. In the fitting results, the minimum of parameter H was 2.06 for deep-focus earthquakes and the maximum of it was 2.62 for intermediate-focus earthquakes. The H values for the three kinds of earthquakes with different depths deviated from that of the whole earthquake data up to -0.38, and the relative deviation was about $15.6\%$. The amplitude R of the spectrum index variation reflects the deviation extent from the average level, and by the fitted results we can see that the deviation amplitudes R became larger and larger going from the shallow-focus shocks to intermediate shocks to deep earthquakes.

 The fitted values of phase P for shallow- and intermediate-focus shocks were consistent with that for the whole earthquakes within the range of statistical error, while the fitted phase P for deep-focus earthquakes deviated from the whole earthquake to a great extent. In term of the goodness of fit $\chi^2/ndf$, the structured sinusoidal function deviated from the real data slightly, but their trends showed agreement with each other. We now devote more attention to the fitted parameter C, which reflects the cycle change of the magnitude spectrum indices.

\begin{table*}[t]
\caption{The fit results of the indices of Eq.(2) for data from Fig.\ref{5} and \ref{cycle}; fitted errors by this equation appear in brackets. WE: Whole earthquake data, SF: Shallow-focus shocks, IF: Intermediate-focus shocks, DE: Deep-focus earthquakes.}
\tabcolsep=15pt
\begin{tabular}{cccccc}
\hline
Parameters&H&R&C(years)&P&$\chi^2/ndf$\\
\hline
WE&2.44 (0.01)&0.40 (0.004)&30.98 (0.62)&-0.4 (7.2)&4.2\\
\hline
SF&2.43 (0.01)&0.40 (0.002)&30.01 (0.57)&1.5 (8.1)&3.4\\
\hline
IF&	2.62 (0.05)	&0.52 (0.06)&	29.98 (0.11)	&2.1 (1.4)	&1.9\\
\hline
DF&	2.06 (0.08)&	1.24 (0.10)&	42.20 (3.11)&	36.1 (21.6)&	1.3\\
\hline
\end{tabular}
\end{table*}

We can see from the results in Table 1 that the cycle of magnitude spectrum indices for shallow-focus shocks was 30.01 years, which approximately approaches the cycle value of 30.98 years for the whole earthquake data set. This was due to the high proportion, i.e., $72\%$, of shallow-focus shocks in the whole data set and the fact that the spectrum indices for shallow-focus shocks played a major role in the results for the whole data set. If we take statistical errors into account, the spectrum index cycles for shallow- and intermediate-focus shocks can be considered to be consistent with each other approximately, and the values were 30.01 and 29.98 years, respectively. Additionally, the range of spectrum indices, average value H, and phase P were also similar for the two samples. Statistically, this may represent possible evidence that the shallow- and intermediate-focus shocks are modulated by a similar triggering dynamical mechanism.

What is a little interesting is that the cycle of the magnitude spectrum for deep-focus earthquakes was around 42.20 years. Why does it show a significant difference with the periodicity of shallow- and intermediate-focus shocks? This will require more rigorous analyses and further theoretical studies of earthquake dynamics and plate tectonics to explain. In term of the index range, the upper limit of spectrum indices for deep-focus earthquakes approached that for shallow- and intermediate-focus shocks, but the lower limit of spectrum indices for deep-focus earthquakes was below that for the shallow- and intermediate-focus shocks. Hence we can deduce that there may be a certain intrinsic relation that exists among the shallow-, intermediate- and deep-focus earthquakes. Additionally, the proportion of large magnitude deep-focus earthquakes was highest at the lowest point of the sinusoidal function cycle.

\subsection{The evolvement of the annual equivalent total magnitude and average epicenter depth}

\begin{figure*}[t]
\includegraphics[width=10cm]{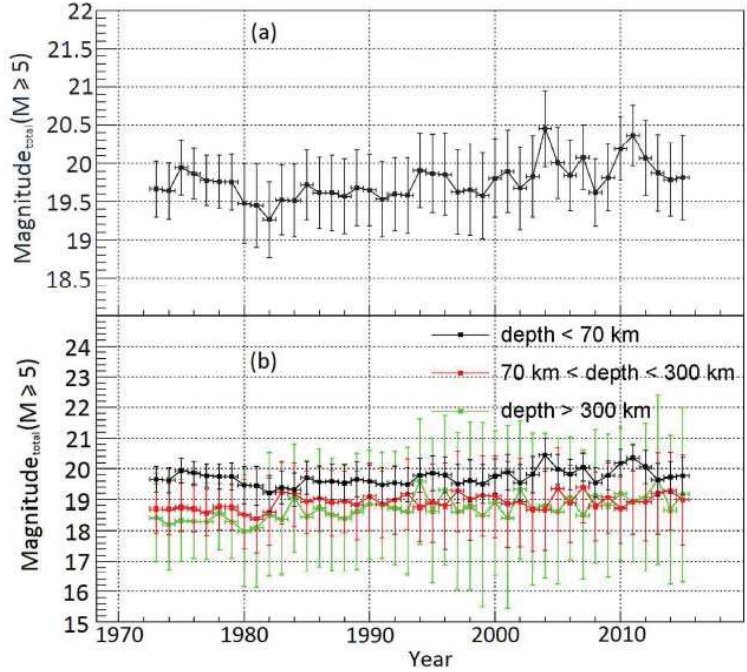}
\caption{The evolution curve of the AETM of the earthquakes ($M\ge5$) for each year.
(a): total earthquake data, (b): different epicenter depth level.
}
\label{aetm}
\end{figure*}

In order to describe the total earthquake energy with $M\ge5$ for one whole year, we put forward the definition of annual equivalent total magnitude (AETM). Because earthquakes occur many times each year and we need to add all the energy emitted by all the earthquakes to analyze the total emitted energy change when comparing them in different years, a new technique had to be devised. The Richter magnitude scale cannot be directly added, so we transformed the Richter magnitude scale into energy units, i.e., joules and then added the joules as the total energy emitted in one year. However since the joule is too large to be used as the final calculation unit for earthquake energy, ultimately we transformed the total energy emitted in one year back into the Richter magnitude scale by use of a transformation function, and the finals values were named as annual equivalent total magnitudes; these data are shown in Fig.\ref{aetm} for Jan. 1973 to Dec. 2015.

From Fig. \ref{aetm}, regardless of whether one views the whole earthquakes or different deep shocks, the average AETM showed an upward tendency with the years, and it changed from 19.3 in 1981 to 20.3 in around 2011. Comparing the plots in Fig. \ref{aetm} , we can see that the energy of the AETM came mainly from the contribution of shallow-focus shocks, except for in 1983 and 1984. The level of the AETM for intermediate- and deep-focus earthquakes was similar.

\begin{figure*}[t]
\includegraphics[width=10cm]{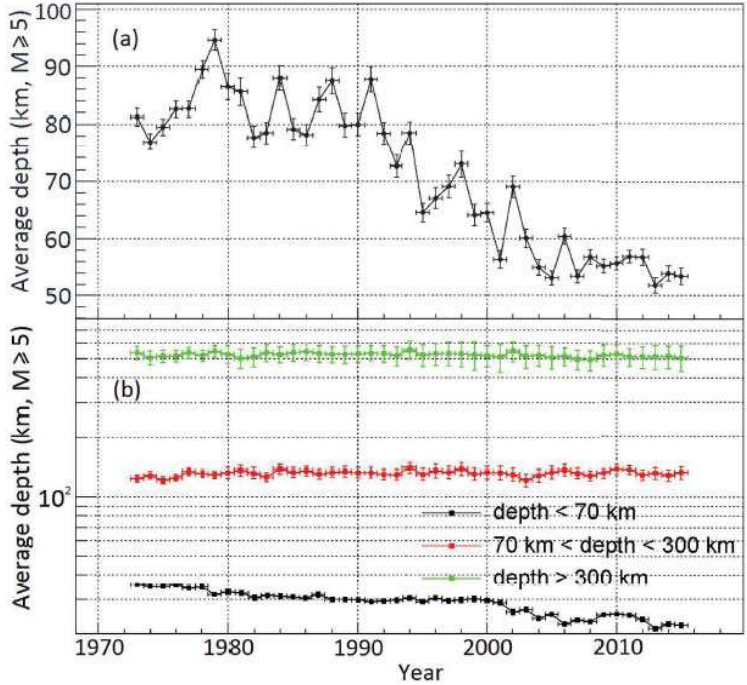}
\caption{The evolution curve of the average epicenter depth of the earthquakes ($M\ge5$) per year.
(a): total data, (b): different depth level.
}
\label{dep}
\end{figure*}

\begin{figure*}[t]
\includegraphics[width=10cm]{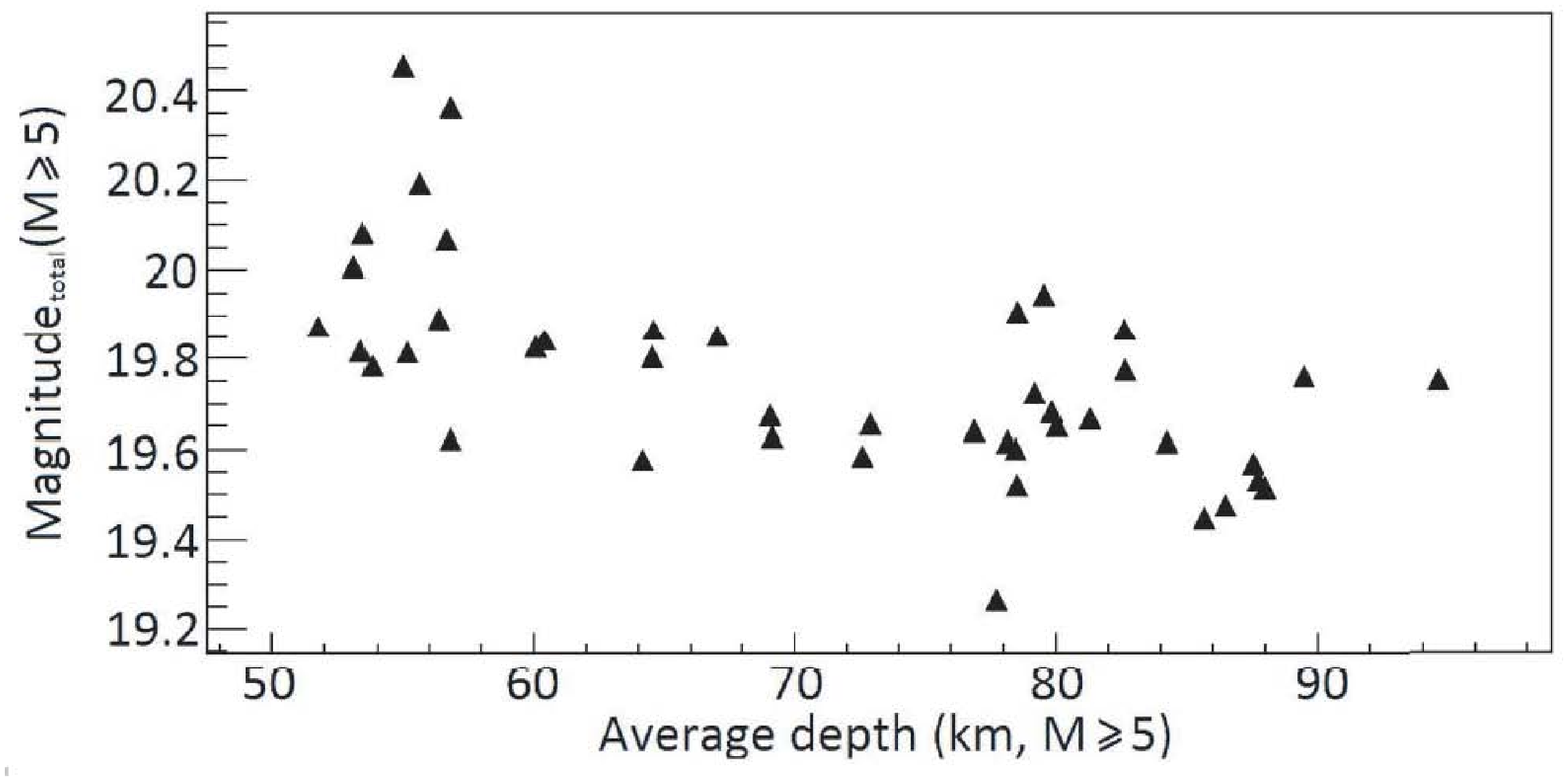}
\caption{The relation between the AETM and the average epicenter depth of the earthquakes ($M\ge5$) for each year. C.C.=-0.61.
}
\label{relat}
\end{figure*}

At the same time, we performed statistical analyses to determine the average epicenter depth of annual earthquakes with $M\ge5$, as shown in Fig. \ref{dep}. From this figure, we can see that from 1973 to now, the average epicenter depths displayed a tendency of becoming shallower and shallower with oscillations, and depths changed from 90 km at the beginning to 50 km afterwards. In the plot (b) in Fig.\ref{dep}, the shallow-focus shocks played a major role in this trend, and depths changed from 35 km at the beginning to 22 km afterwards. The average epicenter depths for intermediate- and deep-focus earthquakes remained basically stable over the years. In the plot (a) in Fig.\ref{dep}, the evolvement of the average epicenter depth of annual earthquakes did not take on a obvious linear distribution, but instead displayed an approximate sinusoidal function.
As shown in Figs. \ref{aetm} and Fig\ref{dep}, the AETM and average epicenter depth showed the opposite trends of evolvement with years. We studied the correlation between the two up plots in Figs. \ref{aetm} and Fig\ref{dep}, i.e. for the total earthquake data, and found that the correlation coefficient was about -0.61, as shown in Fig. \ref{relat}. Obviously, the data were inversely correlated. But only considering the intermediate- and deep-focus shocks data set,  the correlation coefficient between their AETM and average epicenter depths was about zero. Thus the inverted correlation with a coefficient of -0.61 came mainly from the contribution of shallow-focus shocks. In summary, within the time range of 42 years, the shallower the average epicenter depths were, the more annual total energy was emitted , that is, the higher the AETM was. This is a key factor for people to take into account when reinforcing buildings to protect them from earthquakes.

\section{Conclusions and Discussions}\label{conclusion}
Previously published papers about the cyclical characteristics of earthquakes have generally focused on certain local regions and the cyclical characteristics have usually shown direct links with the geologic configuration. The stress accumulation and the emission time for some certain geological configurations is relatively stable and takes on strong period features. Apparently in terms of the frequency of earthquake occurrence shown in Fig.\ref{2}, there are not very many strong period feature for the earthquakes all over the world.

The Earthquake spectrum index directly reflects whether the strong shocks or the weak ones are dominant. The higher the index value is, the larger the proportion of strong shocks; inversely, the lower the index value is, the smaller the proportion of strong shocks. For the first time, we have used the spectrum index fitting method to statistically analyze the magnitude spectrum and duration change regulation with $M\ge5$ earthquakes all over the world. We found that there are obvious cyclic characteristics for the magnitude spectrum distribution.

\begin{figure*}[t]
\includegraphics[width=7cm]{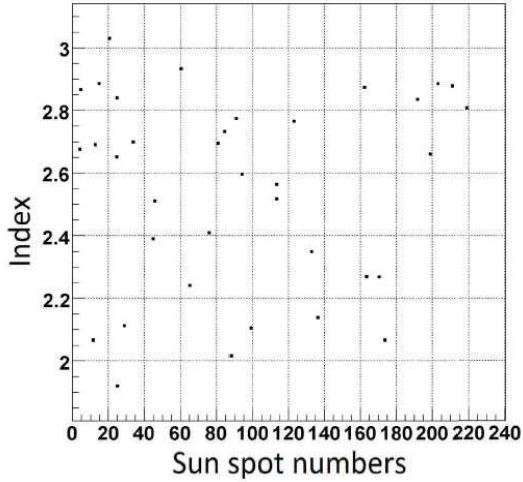}
\caption{The correlation between the fitted index of earthquake magnitude spectrum and the numbers of Sun spots in each years.}
\label{EQdis}
\end{figure*}

On the one hand, there are not any celestial bodies that have a similar cycle like that of the seismic magnitude spectrum indices of 30.98 years and 42.20 years.
 In addition, we studied the correlation between the two spectrum index cycles and the number of sun spots for each year and the correlation coefficient was about zero (data shown in Fig. \ref{EQdis}). This indicates that the 11 year cycle of Sun activity was not the direct inducing mechanism for the cyclic characteristics of the two spectrum indices. However, we have to emphasize in particular that it is a largely different correlation with that between the global earthquake rate and the number of sun spots, that is, the former focus is on the spectrum index of the power exponential distribution of earthquakes, but the later focus is on the earthquake rate.  Rajesh {\it et al.}\cite{Rajesh} applied new techniques and showed that the normalized global earthquake numbers reconstructed from the periodic components of 11 year periodicities had a significant negative correlation (-0.4) with the sunspot number. In recent studies, there are still other works showing obvious correlations between sunspot numbers and earthquakes, and also those showing the centennial scale solar activity effects on seismicity \cite{Shestopalov}. The spectrum index of the power exponential distribution of earthquakes represents the proportion relationship of large magnitude earthquakes and small earthquakes, so this disagreement of the two correlations indicates that the physical mechanism driving the proportion relationship of large magnitude earthquakes may be different from that driving the global earthquake occurrence rate.
We think that the spectrum index cycles of 30.98 years and 42.20 years may possibly originate from motion impacts in the inner Earth, such as mantle movement. They may also probably be attributed to the lagging effect caused by the comprehensive cyclic impact on the inner Earth involving Earth's tropical year cycle, Earth's rotational cycle, the lunar orbital period and the orbit of the moon, and the intersection of the precession cycle.

 Moreover, phenomenons whereby the AETM becomes higher and higher and the average epicenter depth becomes more shallow occur at the same time. Objectively, the average epicenter depth cannot possibly continue to become shallower up to zero in the future, so there may be certain cyclic characteristics that exist over a longer time scale. 
 The fact that AETM and the average epicenter depth took on a strong negative correlation indicates that they have a close relation with each other, but the precise physical mechanism will require more rigorous data analyses using more data accumulated over a more accurate and longer earthquake record to study further and confirm.

\section{Acknowledgement}
We are appreciative of the USGS website that provided the earthquake data. We acknowledge the National Science Foundation (11103023), which partially supported this work. Additionally we are grateful to Prof. Y. Q. Ma for her valuable and instructive discussions regarding this work.



\renewcommand\refname{Reference}

\bibliographystyle{h-physrev}

\bibliography{SinEQ}

\end{document}